\documentclass[acmsmall,screen]{acmart}

\usepackage{graphicx}
\usepackage{tabularx}
\usepackage{booktabs} 
\usepackage{color}
\usepackage{multirow}
\usepackage{bm}
\usepackage{algpseudocode}
\usepackage[ruled,linesnumbered]{algorithm2e}

\AtBeginDocument{%
  }

\setcopyright{none}

\acmJournal{TIST}
\acmVolume{00}
\acmNumber{0}
\acmArticle{000}
\acmMonth{0}

\begin{document}

\title{Explicit Knowledge Graph Reasoning for Conversational Recommendation}

\author{Xuhui Ren}
\email{x.ren@uq.edu.au}
\affiliation{%
  \institution{The University of Queensland}
  \city{Brisbane}
  \state{Queensland}
  \country{Australia}
  \postcode{4072}
}

\author{Tong Chen}
\email{tong.chen@uq.edu.au}
\affiliation{%
  \institution{The University of Queensland}
  \city{Brisbane}
  \state{Queensland}
  \country{Australia}
  \postcode{4072}
}

\author{Quoc Viet Hung Nguyen}
\email{henry.nguyen@griffith.edu.au}
\affiliation{%
  \institution{Griffith University}
  \city{Gold Coast}
  \state{Queensland}
  \country{Australia}
  \postcode{4215}
}

\author{Lizhen Cui}
\email{clz@sdu.edu.cn}
\affiliation{%
  \institution{Shandong University}
  \city{Jinan}
  \state{Shandong}
  \country{China}
  \postcode{250100}
}

\author{Zi Huang}
\email{huang@itee.uq.edu.au}
\affiliation{%
  \institution{The University of Queensland}
  \city{Brisbane}
  \state{Queensland}
  \country{Australia}
  \postcode{4072}
}

\author{Hongzhi Yin}
\authornote{Corresponding author.}
\email{h.yin1@uq.edu.au}
\affiliation{%
  \institution{The University of Queensland}
  \city{Brisbane}
  \state{Queensland}
  \country{Australia}
  \postcode{4072}
}

\renewcommand{\shortauthors}{Ren et al.}

\begin{abstract}
  Traditional recommender systems estimate user preference on items purely based on historical interaction records, thus failing to capture fine-grained yet dynamic user interests and letting users receive recommendation only passively. Recent conversational recommender systems (CRSs) tackle those limitations by enabling recommender systems to interact with the user to obtain her/his current preference through a sequence of clarifying questions. Despite the progress achieved in CRSs, existing solutions are far from satisfaction in the following two aspects: 1) current CRSs usually require each user to answer a quantity of clarifying questions before reaching the final recommendation, which harms the user experience; 2) there is a semantic gap between the learned representations of explicitly mentioned attributes and items.

To address these drawbacks, we introduce the knowledge graph (KG) as the auxiliary information for comprehending and reasoning a user's preference, and propose a new CRS framework, namely \textbf{K}nowledge \textbf{E}nhanced \textbf{C}onversational \textbf{R}easoning (KECR) system. As a user can reflect her/his preference via both attribute- and item-level expressions, KECR closes the semantic gap between two levels by embedding the structured knowledge in the KG. Meanwhile, KECR utilizes the connectivity within the KG to conduct explicit reasoning of the user demand, making the model less dependent on the user's feedback to clarifying questions. KECR can find a prominent reasoning chain to make the recommendation explainable and more rationale, as well as smoothen the conversation process, leading to better user experience and conversational recommendation accuracy. Extensive experiments on two real-world datasets demonstrate our approach's superiority over state-of-the-art baselines in both automatic evaluations and human judgments.
\end{abstract}



\keywords{conversational recommendation, knowledge graph, preference mining}


\maketitle

\section{Introduction}
Recently, conversational recommender systems (CRSs) have attracted widespread attention from both academia and industry for providing persuasive and explainable recommendations through human-like conversations with users \cite{10.11453394592,DBLPabs181207617, liuconversational, xuetal2020user}. Different from conventional recommender systems that produce recommendations by analyzing the user preference from their historical interactions, CRSs proactively ask questions to help users explicitly clarify their up-to-date demand during the conversation, achieving dynamic and explainable recommendations.

The general framework for a CRS consists of a dialogue component (DC) and a recommender component (RC) \cite{journalscor09459}. The responsibility of the DC is to generate clarifying questions and replies to further the conversation. The clarifying questions guide a user to express preference on the specific attribute types in a ``System Ask-User Answer'' (SAUA) manner \cite{1011453269206271776}. RC extracts the contextual information inside the conversation utterances and recommends suitable items to satisfy the user's preferences when the confidence is high. Recently, a variety of CRSs have been proposed to improve interactions with users and pursue high-quality recommendation \cite{10.11453394592, DBLPabs181207617, liuconversational, xuetal2020user, 1011453269206271776, 10114533944863403143}.

Despite that existing CRSs have demonstrated improvements over conventional recommender systems, two intrinsic issues of CRSs have been hindering their practicality. First, current CRSs heavily rely on the user's responses to the clarifying questions to acquire timely attribute-level demands. However, the considerably large attribute pool in real-world applications leads to numerous system attempts on formulating clarifying questions, which weakens the user experience \cite{lin2023enhancing} as a user will be required to answer a long list of questions before receiving the final recommendation \cite{1011453404833462839}. Second, such an SAUA interaction scheme easily triggers a semantic gap between the explicit attribute-level and item-level user preferences \cite{10114533944863403143}. While it is common for a user to express her/his preference at both levels, the SAUA mode can only capture the superficial keywords appearing in the utterances, lacking complementary contextual information for fusing attribute- and item-level expressions to better understand user interests. As shown in Table \ref{tab:CRS_example}, a user is asking for scary movies like ``Annabelle'', where her/his demand is expressed by attribute-level expression ``horror'' and item-level expression ``Annabelle''. Ideally, if the CRS could make full use of utterance contexts, it may naturally deduce a persuading recommendation with a reasonable explanation, e.g., ``a horror film and a look into the early works of James Wan'' in the example. However, the straightforward keyword extraction in contemporary CRSs cannot support the comprehension of underlying semantics that bridge ``horror'' and ``Annabelle'', not to mention fusing them to generate accurate recommendations. Therefore, an effective solution is needed for filling the semantic gap when understanding user utterances.

\begin{table}[]
	\caption{An example of a conversation for movie recommendation. Mentioned items and attributes are marked in blue and red, respectively.}
	\vspace{-0.3cm}
		\begin{center}
			\begin{tabular}{|l|p{0.8\columnwidth}|}
				\hline
				User:	&Hi, I am looking for a movie recommendation.  \\ 
				System:	&What kind of movies do you like?  \\
				User:	&  I love \textcolor{red}{horror} movies similar to \textcolor{blue}{Annabelle}. I never knew a doll could be so scary.\\ 
				System:	&  \textcolor{blue}{Dead Silence} might be suitable for you! It is a \textcolor{red}{horror} film and a look into the early works of \textcolor{red}{James Wan}.  \\ 
				User:	&  Really? I would like to have a try! \\ \hline
			\end{tabular}
		\end{center}
		\bigskip\centering		
	\label{tab:CRS_example}
	\vspace{-0.5cm}
\end{table}

Some recent CRSs incorporate auxiliary knowledge graphs (KGs) as complementary resources for analyzing item- and attribute-level user preferences \cite{chental2019towards, 10114340483247, sarkaretal2020suggest,zhang2023variational}. However, existing work merely utilizes the graph-structured knowledge in an implicit fashion, i.e., the knowledge is distilled via approaches like graph neural networks \cite{101007978-393417438} into a latent representation of user preference, which voids a KG's advantage in explicit reasoning. In fact, the inherent graph structure of the KG could naturally link the item- and attribute-level preferences to facilitate explicit reasoning of the user demand \cite{chental2019towards, jietal2020language}, providing a more efficient way to engage the DC and the RC. From the perspective of the DC, the reasoning process could enable smoother response generation with stronger rationality. The dialogue policy can follow the reasoning flow to proactively explore user interests, chat by referring to informative entities, and provide recommendations with faithful explanations \cite{moonetal201pendialkg}. From the perspective of the RC, reasoning over the KG uncovers the connections between users and items, thus helping to fully comprehend a user's preferences. The extra connectivity information presents chains of evidence to support explainable recommendation, and is also beneficial for tackling the cold-start problem during recommendation \cite{10114556437564421, WangWangXueCaoChua2019}.

In this paper, we propose a \textbf{K}nowledge \textbf{E}nhanced \textbf{C}onversational \textbf{R}easoning (KECR) approach for CRS that conducts explicit reasoning over the KG. The key hypothesis of this work is that an explicit reasoning process on the KG can better stimulate the strengths of a CRS, making it rational and explainable. An accurate and rational graph reasoning process usually requires to learn the entity embeddings and the context embeddings. Therefore, we incorporate an item-based KG (i.e., DBpedia \cite{bizer2009dbpedia}) to model the semantic embeddings of the item/attribute entities, and a pre-trained language model to learn the context embeddings. As these two embeddings are from two different embedding spaces, we additionally utilize mutual information estimation \cite{hjelm2018learning} to optimize embeddings by forcing the token embeddings in two embedding spaces to be closer given the utterance-entity co-occurrences in the conversations. The optimized embeddings encode comprehensive context information and user preferences in a recommendation setting, and constitute the inputs of the reasoning module. The reasoning module makes entity selection over the KG to find a prominent path from the mentioned items and/or attributes to a suitable recommendation, constructing a reasoning flow to guide the system to generate responses. Supervised by this explicit graph reasoning fashion, KECR selects the most relevant entities in the reasoning flow to generate human-like responses with a pre-trained language model, which could guarantee the coherence of the conversation content and improve the recommendation interpretability \cite{wang2022towards}.

To the best of our knowledge, few CRS methods could explicitly reason over the KG to find suitable recommendations as well as explanations. Our proposed KECR simultaneously takes advantages of both the structured knowledge to bridge the semantic gap between the item-level and attribute-level preference expressions, and leverages the graph connectivity to infer a chain of evidence to support explainable recommendations. Finally, we conduct extensive experiments on two real-world datasets to verify the effectiveness of our approach regarding both the recommendation accuracy and the quality of the generated conversations.

\section{Related Work}
\subsection{Knowledge Graph Reasoning}
Performing explicit reasoning on a knowledge graph (KG) has been demonstrated as an effective solution to retrieving a user's desired items or answers \cite{qiuetal201dynamically, fengtal20scalable}. Prior efforts are focused on multi-hop reasoning over KGs to support question-answering tasks. The queries are augmented with entity graphs or concept nets to enhance their semantic representations, which are then fed into the database to facilitate similarity-based retrieval \cite{xiongetalimproving, sunetal2019ullnet}. Motivated by the superior performance on item prediction, some recommender systems leverage KG embedding techniques to improve item embeddings during the recommendation \cite{1011453292500330991, chental2019towards, 10114533944863403143, 1011453404833462839}. However, such methods cannot explicitly utilize the connection information within the graph, resulting in the loss of explainability.

To address this problem, several path-based methods are proposed. The path-based methods are designed to explore prominent paths between users and recommended items with the connection of entities in the knowledge graph \cite{ 101153219813219965, 101145357384357805}. However, they lack an interactive approach to obtain users' needs, and their reasoning processes are limited to pre-defined meta paths and cannot be dynamically refined with users' new inputs \cite{WangWangXueCaoChua2019}. Consequently, their performance heavily relies on the additional domain knowledge for meta-path engineering, which is rather labour-intensive.

\begin{figure}
	\centering
	\includegraphics[width=0.85\linewidth]{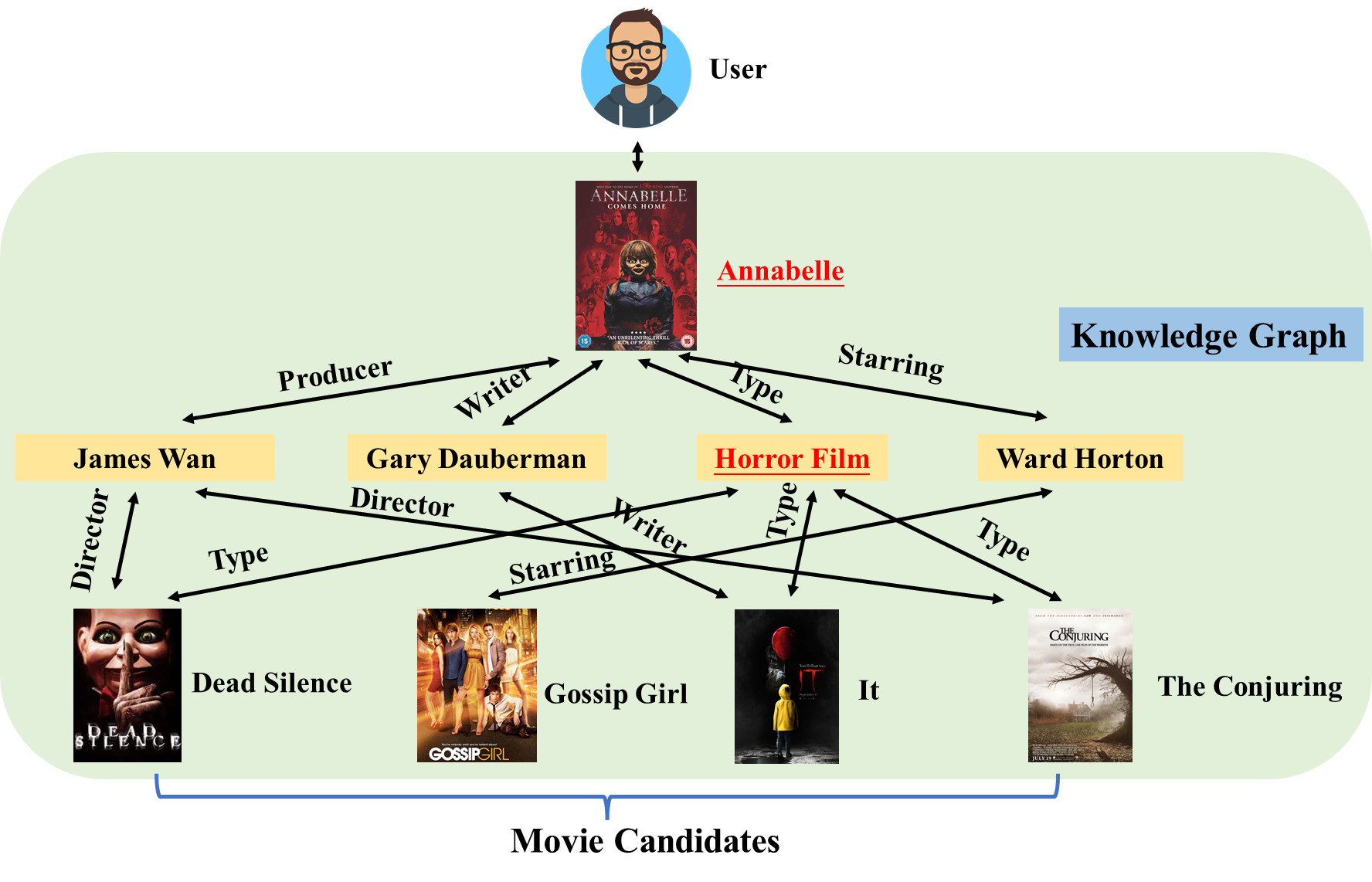}
	\vspace{-0.5cm}
	\caption{An illustration of connection information inside the KG in the movie domain. The solid lines between entities represent their relations, while the underlined entity marked in red represents what the user mentioned during the conversation.}
	\label{fig:kgsample}
	\vspace{-0.5cm}
\end{figure}

\subsection{Conversational Recommendation}
CRSs have become an appealing approach to bridge the interactions between users and systems for meeting users' real-time demand with explainable and accurate recommendations. This emerging research topic originates from task-oriented dialogue systems that help users complete recommendation-related tasks in a multi-turn conversation \cite{1032099783210002, 1011453404833462839}. The conventional design of CRSs follows the SAUA fashion \cite{1011453269206271776}, where the DC asks a series of clarifying questions over pre-defined attributes to collect the user's explicit preference and use a retrieval/collaborative filtering-based (CF) recommendation method \cite{hu2022learning} to deduce personalized recommendations. The main limitation of this kind of methods is that, real-world application scenarios contain a large number of candidate attributes that can be used to formulate clarifying questions. As a result, a user may be required to answer many clarifying questions before reaching the final recommendation, which greatly harms the user experience.
Also, the dialogue behaviours and content in these methods are strictly restricted. The user and the CRS can only interact with the user's preference on a fixed set of attributes, and the CRS straightforwardly presents recommendations without any explanations \cite{DBLPabs201010333}. It thus leads to the semantic gap in the utterance understanding, especially when the user reflects his/her preference towards both items and attributes.

Some recent approaches adopt knowledge graphs (KGs) to enrich the conversation information to bridge the semantic gap during the understanding \cite{wang2022towards,chental2019towards, 10114340483247, sarkaretal2020suggest, jietal2020language, wenqiangleikdd20}. For example, an item attribute graph is leveraged in \cite{tu2022conversational,zhou2022cr} to effectively bind the hierarchical information with the recommendation and conversation generation process. As discussed in Section 1, most existing works along this line do not make full use of the advantages of the knowledge graph (e.g., learning an implicit representation of the subgraph retrieved from the KG \cite{zhang2023variational}). We argue the connectivity information over the knowledge graph is a crucial piece of information for comprehending users' preferences. Explicitly reasoning over the knowledge graph would better carry forward the advantages of CRSs. To this end, we devise a novel conversational recommender system by explicitly reasoning the user preference on the knowledge graph. As such, our model is able to efficiently and precisely learn user preferences and provide reasonable guidance for response generation.

\section{Preliminaries}
Conversational recommendation is currently an emerging research topic that has aroused exploration in various perspectives \cite{1032099783210002, 10114540508240605, wenqiangleikdd20, 1011453404833462839}. This work follows the basic settings of CRSs \cite{1032099783210002} but makes some changes to improve its real-world applicability. In a conversation, the CRS is free to interact with the user via several coherent utterances to acquire the user's real-time preferences. According to the collected preferences, the system will deduce suitable recommendations to satisfy the user. The entire conversation ends until the recommendation has been accepted or the user leaves. This paper points out that reasoning on the KG better takes advantage of the graph's structural information to generate rational and explainable recommendations. Following the movie recommendation example in Table \ref{tab:CRS_example}, a user wants a \textit{horror} movie that is similar to \textit{Annabelle} she/he has watched. As shown in Figure \ref{fig:kgsample}, such contextual information helps to generate a reasoning path on the knowledge graph to the potential recommendation \textit{Dead Silence} as well as deduce the rationale that the user may prefer this recommendation due to the directing style of \textit{James Wan}. The reasoning process unveils the reason behind a recommended item that distinguishes it from other candidates. So, we incorporate a reasoning function into the CRS setting to endow it with explainability. Simultaneously, the reasoning path can be utilized by the DC to generate sensible persuasions, making the system dialogues more appealing and acceptable.

Without loss of generality, we denote a conversation consisting of several interaction rounds by $\mathcal{X}=\{x_t\}_{t=1}^{\phi}$, in which each interaction round $x_t$ refers to a pair of response utterances between the system and a user and $\phi$ is the number of conversation rounds. To facilitate the reasoning process, we resort to an auxiliary knowledge graph $\mathcal{G}= (\mathcal{V}, \mathcal{R})$, where $\mathcal{V}$ denotes the entity set and $\mathcal{R}$ denotes the relation set. Both items and their associated attributes are represented as nodes (a.k.a. entities) in the KG. Each item entity is linked with its corresponding attribute entities via bidirectional relations, e.g, $\textnormal{\textit{Annabelle}} \stackrel{Type}{\Longleftrightarrow} Horror \,Film$. In each conversation round $t$, the system selects a subset of neighbourhood entities $v\in \mathcal{V}$ to enhance the response generation, which could be regarded as a reasoning propagation process over the KG. The purpose is to optimize the reasoning path to maintain the engagement and semantic rationality of the conversation, and output the response utterances and eventually make an item recommendation to the user.

\section{Methodology}
\begin{figure*}[!t]
	\centering
	\includegraphics[width=0.9\linewidth]{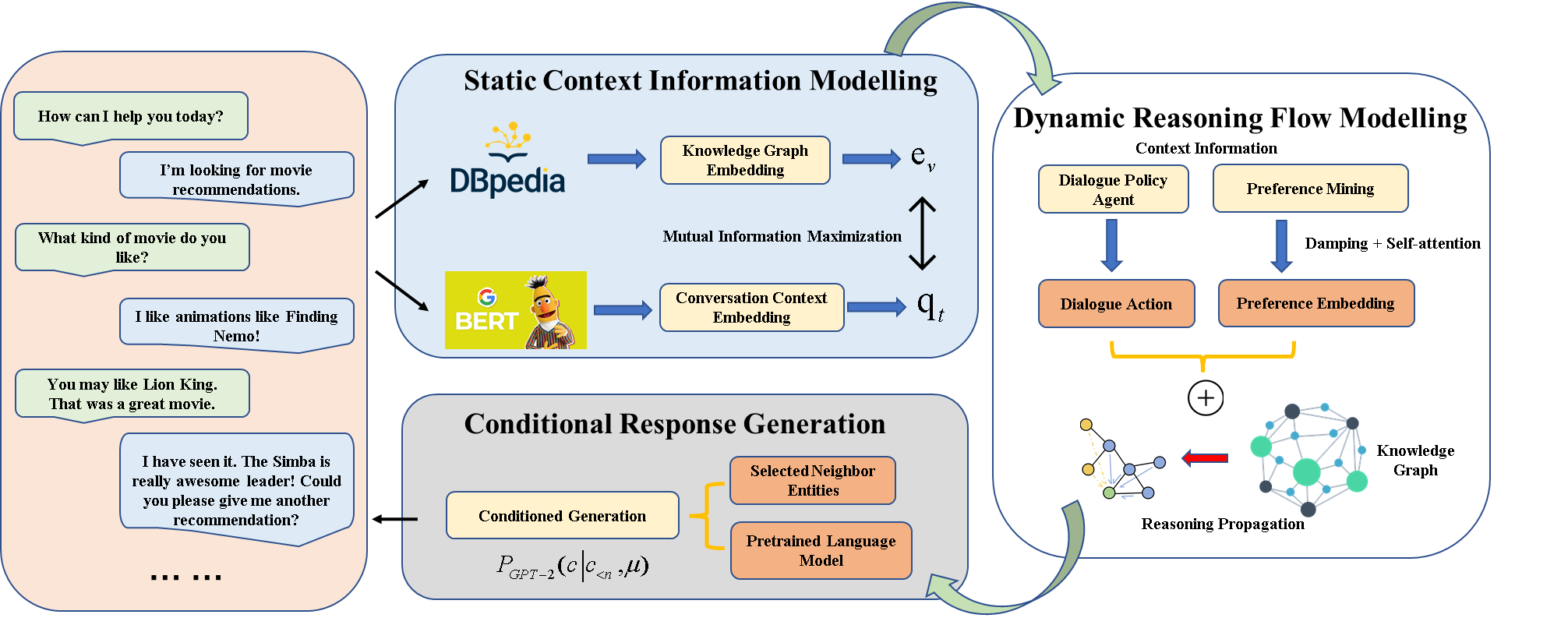}
	\caption{The overall workflow of KECR. KECR is composed by three key components: Static Context Information Modelling, Dynamic Reasoning Flow Modelling and Conditional Response Generation.}
	\label{fig:framework}
	\vspace{-0.5cm}
\end{figure*}

We describe our proposed Knowledge Enhanced Conversational Reasoning (KECR) approach in this section. As shown in Figure \ref{fig:framework}, KECR is mainly consisted of three components: (1) Static Context Information Modelling that is designed for embedding the static utterances and the KG into latent representations; (2) Dynamic Reasoning Flow Modelling that extracts the instant context information to estimate the user preference and predict the next response action, and then reason over the KG to select proper entities for the downstream task; and (3) Conditional Response Generation that produces human-like responses based on the selected entities. The design of each component is detailed in the following subsections.

\subsection{Static Context Information Modelling}
As illustrated in Section 1, the diversified expressions of user preferences (i.e., item- and attribute-level expressions) make it difficult to understand the underlying interests within a user utterance, hindering the CRS from deducing accurate recommendations and suitable dialogue responses. We propose to tackle this problem from two perspectives in this paper to improve the learning of such context information. Firstly, we introduce a KG to enrich the semantics of a user's expression, thus uncovering the associations between item- and attribute-level user preferences. Then, a pre-trained language model is adopted to learn the semantic representation of the conversation context.

\subsubsection{Knowledge Graph Embedding}
To build a KG for our CRS, we follow \cite{chental2019towards, 10114533944863403143} to reconstruct a subgraph from DBpedia that only contains item entities appearing in our corpus. The original DBpedia only contains directional links from items to their corresponding attributes, which is insufficient for performing embedding and reasoning over the KG in the recommendation context. Therefore, we expand the item-attribute relation to be bidirectional, and add new links from generic categories (e.g., movie genre and release date) to their values, compensating the information within the KG.

As the connected neighbours of an entity could reflect its inherent properties, recent methods based on graph convolutional networks (GCNs) have shown state-of-the-art effectiveness in aggregating neighborhood data to enhance the learned entity embeddings of a KG \cite{Zhanhuahe020, xiongetaldeeppath, 103292500330989}. For modelling this relational information, we adopt R-GCN \cite{1031993417438}, which generalizes GCN with relation-specific weight matrices, to learn entity representations in a finer granularity. At the $(l\!+\!1)$-th layer, the embedding of each entity $v\in \mathcal{V}$ is learned by propagating information from its one-hop neighbours:
\begin{equation}\label{eq:NRN}
\mathbf{e}_v^{l+1}=\sigma(\sum_{r \in R} \sum_{v^{'} \in \mathcal{N}_r(v)} \frac{1}{z_v^r} \mathbf{W}_r^l \mathbf{e}_{v^{'}}^l +\mathbf{W}_0^l \mathbf{e}_v^l),
\end{equation}
where $\sigma$ is the sigmoid function, $\mathcal{N}_r(v)$ denotes the neighbor entity set of $v$ under relation $r\in \mathcal{R}$, $z_{v}^{r}$ is the normalization factor, $\mathbf{W}_r^l$ and $\mathbf{W}_0^l$ are learnable weight matrices for integrating the relation-specific neighbors and self-loop at the $l$-th layer, respectively. For notation convenience, we denote the entity embedding at the final layer as $\textbf{e}_v$, which is used for subsequent computations.

\subsubsection{Conversation Context Embedding}
During a conversational recommendation process, the system interacts with the user to acquire necessary information about the real-time user preference from the conversation context and generate high-quality recommendations. The learning of the context embedding influences the a CRS's comprehension of the user preference. Recent efforts in adopting a large pre-trained language model (e.g., BERT \cite{devlinetal201t} and GPT-2 \cite{budzianowvu9ello}) in an open-domain dialogue system have achieved substantial improvements on modelling the conversational context \cite{zhaneta01rsonalizing}, showcasing the potential of exploiting large pre-trained language models in conversational recommendation.

In this vein, we adopt BERT, a pre-trained multi-layer transformer to embed the conversation utterances. In each conversation round, the concatenation of the conversation utterances between the system and the user in $x_t\in \mathcal{X}$ is fed into BERT (as a sequence of word IDs) to obtain its latent embedding $\mathbf{x}_t$. To enhance the contextual connections with the conversation history, we build a gated recurrent unit (GRU) \cite{cho14earning} to account for long-term sequential dependencies. The context embedding in round $t$ is obtained via:
\begin{equation}\label{eq:GRU}
\mathbf{q}_t=GRU(\mathbf{q}_{t-1}, \mathbf{x}_t).
\end{equation}
In addition to the context embedding $\mathbf{q}_t$, the entities mentioned in the conversation are in particular important for understanding the user's preference. To recognize these entities during the conversation, we perform fuzzy matching with the lemmatized form of user utterances using Spacy\footnote{https://spacy.io/} and map them into the KG. In round $t$, we store the KG embeddings of entities in all previous $t$ rounds as a dialogue belief state, denoted by $\mathbf{D}_t = [\textbf{e}_{v_1}, \textbf{e}_{v_2},\cdots,\textbf{e}_{v_E}]$ with $E$ chronologically sorted entities.

\subsubsection{Learning Embeddings via Mutual Information Estimation}
We hereby describe how we jointly learn both KG and conversation context embeddings. Given the constructed KG and full conversation $\mathcal{X}$, one may choose to learn each type of the embeddings separately. However, this aggravates semantic discrepancies as the representations of the KG entities and conversation context are generated from two distinct embedding spaces. In an utterance, as the mentioned entities $v$ are informative and heavily entangled with the conversation context, we take advantage of the mutual information (MI) maximization trick \cite{c921e40702284b7,yehainfomax,10114533944863403143} to align and fuse highly relevant KG and conversation embeddings.
The main idea is to calculate the MI between two coupled variables, i.e., the co-occurring entity $v\in \mathcal{V}$ and utterance $x_t\in \mathcal{X}$ in our case:
\begin{equation}\label{eq:KL}
MI(v,x_t) = D_{KL}(p(v,x_t) \rVert p(v)p(x_t)),
\end{equation}
where $D_{KL}$ is the Kullback-Leibler divergence between the joint distribution $p(v,x_t)$ and product of marginals $p(v)p(x_t)$.

Given the intractable computation of the precise MI in practice, we adopt a neural approach to estimate MI using Jensen-Shannon divergence \cite{hjelm2018learning}, which could be implemented in the following binary cross-entropy (BCE) form $L_{MI}$:
\begin{equation}
\begin{split}
L_{MI} &= \mathbb{E}_P[\log(g(\mathbf{e}_{v^+},\mathbf{q}_t))] + \mathbb{E}_N[\log(1-g(\mathbf{e}_{v^-},\mathbf{q}_t))] \\
&\leq MI(v,x_t)
\end{split}
\end{equation}
where $\mathbb{E}_P$ and $\mathbb{E}_N$ denotes the expectation over all positive and negative samples, respectively. $g(\cdot,\cdot)$ is a neural classifier, which inputs the concatenation of the embedding vectors of one entity and utterance and computes a probability in the range $[0,1]$. In KECR, for every positive sample $(v^+, x_t)$ consisting of a co-occurring utterance and a mentioned entity, we pair it with four negative samples $(v^-, x_t)$ by randomly drawing entities that do not appear in utterance $x_t$.

By maximizing $L_{MI}$, we can essentially push up the lower bound of MI. In this static context information modelling process, it helps us effectively pre-train both the KG embeddings and the GRU encoder in Eq.(\ref{eq:GRU}) by calibrating and fusing their mutual information. This brings more expressiveness to $\textbf{e}_v$ and $\textbf{q}_t$ generated, which will be used and further finetuned for subsequent reasoning and recommendation tasks.

\subsection{Dynamic Reasoning Flow Modelling}
Conditioned on the static information extracted from the KG and utterance, we need to dynamically comprehend the real-time user preference and reason over the KG to find proper entities to push forward the conversation and/or generate an accurate recommendation. There are mainly two factors that influence the reasoning process, namely the dialogue action the model is about to execute (e.g., query or recommend) and the user preference mined from the conversation. The generated dialogue action will help rectify the reasoning flow to be coherent with the conversation context. Meanwhile, the user preference will help select entities that are the most suitable for the user's demand. In this section, we will detail how both factors are computed and then utilized for explicit reasoning over the KG.

\subsubsection{Dialogue Policy Agent}
The dialogue policy agent decides which dialogue action should be performed conditioned on the current conversation context. Previous CRSs usually define the dialogue action from the perspective of the SAUA-based interaction scheme, where each action is defined as one yes/no question about a specific attribute, e.g., ``Do you like \textit{action} movies?''. As pointed out recently in \cite{1061913371769, 1011453404833462839}, this leads to an unnecessarily large action space for training a performant policy agent, and is less efficient in acquiring user preferences. In this regard, we narrow down the action space to only three types of actions: query, recommend, and chit-chat, while the reasoning process further decide which relevant attribute (i.e., the KG entity) will be incorporated for composing the conversational response.

Inspired by prior works on task-oriented dialogue systems that estimate system actions based on the summarization of the conversation context \cite{DBLPabs201010333, DBLPmWenMBY17}, we train a policy agent with a multi-layer perceptron classifier to decide the dialogue action in each conversation round. As the pre-trained conversation context encoder is sufficiently expressive, we only take the conversation embedding as the input to the policy agent:
\begin{equation}\label{}
\mathbf{a}_t = \mathrm{softmax}(\mathbf{W}_1 \mathrm{ReLU}(\mathbf{W}_2 \mathbf{q}_t +b_2) +b_1),
\end{equation}
where $\mathbf{a}_t$ represents the probability distribution over three dialogue actions, $\mathbf{W}_1, \mathbf{W}_2, b_1, b_2$ parameterize the policy agent with learnable weights and biases. The policy agent is learned by minimizing the following cross-entropy loss:
\begin{equation}\label{}
L_{a} = - \sum_{\forall i}a^*_{t,i} \log(a_{t,i}),
\end{equation}
where $a^*_{t,i}$ and $a_{t,i}$ denote the $i$-th element of the ground truth and prediction, respectively.

\subsubsection{Preference Mining}
As introduced in Section 4.1.2, in each round, the informative entities mentioned in the conversation are stored in $\mathbf{D}_t$. Intuitively, the encodings of these informative entities are strong reflections of the user preference. Instead of simply averaging these vectors, we further consider two marginal effects from the context and time perspectives.

The context-awareness requires to identify the dominant features among the input context rather than treating all entities equally. Therefore, we adopt a simplified self-attention mechanism \cite{DBLPscorrinFSYXZB17} to calculate the context-level user preference representation:
\begin{equation}\label{}
\begin{split}
\mathbf{u}_{cont}= \mathbf{D}_t\cdot \mathrm{softmax}(\mathbf{W}_3 \mathrm{tanh}(\mathbf{W}_4 \mathbf{D}_t))
\end{split}
\end{equation}
where $\mathbf{W}_3$ and $\mathbf{W}_4$ are weight matrices.

The time marginal effect comes from the topic/interest shifts during the conversation. The user may change his/her mind and request for different recommendations within a conversation. Therefore, we adopt a damping factor $\bm{\gamma} = [\gamma_1, \gamma_2,\cdots,\gamma_E]$ to control the intensity of the information influence from the conversation history, and the time-level user preference is formulated as:
\begin{equation}\label{}
\mathbf{u}_{time}=\mathbf{D}_t \bm{\gamma}.
\end{equation}
With that, the final user preference embedding is the mean between $\textbf{u}_{cont}$ and $\textbf{u}_{time}$:
\begin{equation}\label{}
\textbf{u}_t=\frac{\textbf{u}_{cont} + \textbf{u}_{time}}{2}.
\end{equation}

\subsubsection{Heterogeneous Reasoning over the KG}
To perform explicit reasoning on the KG, we devise a heterogeneous graph reasoning module to expand the dialogue action and user preference into a reasoning path to select the right entity for dialogue generation, e.g., choosing an attribute to query about or identifying items to recommend. 

Specifically, the starting point of a reasoning process is the entity $v_{t-1}$ appearing in the previous round of the conversation. If no entity has been mentioned in the user utterance at the early stages of the conversation (e.g., $t=1$), one generic attribute type (e.g., actor, type, genre in movie domain) will be randomly selected to formulate a clarifying question. As the utterances in conversational recommendation tend to be short, few utterances contain multiple entities. If the user does mention multiple entities in a single utterance, one will be randomly selected as the starting point. Though selecting the entity with higher relevance is more desired on such occasions, it requires either the user profile or interaction history to support semantic matching. However, most conversational recommendation scenarios are in strict cold-start conditions, where each conversation is regarded as an individual session without prior knowledge on any user.

With the starting point $v_{t-1}$, the reasoning target is to determine the next entity of the reasoning path from its neighbors $v\in \mathcal{N}(v_{t-1})$, and the next entity should be coherent to the given contexts. Therefore, we need to calculate a relevance score to evaluate the confidence on each neighbour node under the current context, namely the user preference embedding $\textbf{u}_t$, conversation context $\textbf{q}_t$, and dialogue response action $\textbf{a}_t$:
\begin{equation}\label{eq:reasoning}
\begin{split}
\mathcal{J}(k) &= \sigma(\mathbf{h}_k \mathbf{W}_{proj} \mathbf{h}_c),\\
\mathbf{h}_k &= [\mathbf{e}_{v_{t-1}}; \mathbf{e}_k],\\
\mathbf{h}_c &= [\mathbf{a}_t; \mathbf{u}_t; \mathbf{q}_t],
\end{split}
\end{equation}
where $k$ indexes entity $v_k \in \mathcal{N}(v_{t-1})$, $\mathbf{e}_v$ and $\mathbf{e}_k$ are the respectively the KG embedding of the starting and neighbour entities. The relevance score $\mathcal{J}(k)$ denotes each $v_k$'s relevance to the starting point $v_{t-1}$. The neighbour that holds the highest relevance score will be selected in the reasoning path. Since this neighbour  selection process could be regarded as a binary decision on each $v_k \in \mathcal{N}(v_{t-1})$, we optimize this process by minimizing the following BCE loss:
\begin{equation}\label{}
L_{r}^j = -\sum_{v_k \in \mathcal{N}(v_{t-1})} \mathcal{J}(k) \log \mathcal{J}(k) +(1-\mathcal{J}(k))\log (1-\mathcal{J}(k)),
\end{equation}
where the loss is notably indexed by $j$. The rationale is, when a recommendation action is needed and an item entity is reached as the one-hop neighbor of $v_{t-1}$, we expect our system to be able to provide both the recommendation and the reason behind this item selection. Therefore, after selecting the one-hop neighbor of $v_{t-1}$, in KECR we propose to repeat the reasoning process from this selected neighbor entity. The reasoning is performed analogously using Eq.(\ref{eq:reasoning}). As such, it can reason further down the KG path, retrieve attributes closely associated with the item, then formulate an explanation to enhance the rationality and persuasiveness of the system response. At the same time, it also makes the second reasoning step easily impacted by the result of the first step. Hence, we append a small weight $\lambda$ to the second reasoning step $j=2$ to avoid unexpected cumulative error. The final reasoning loss $L_{r}$ is:
\begin{equation}\label{}
L_{r} = L_{r}^1 + \lambda L_{r}^2.
\end{equation}

\subsection{Conditional Response Generation}
The generator integrates selected entities and the context into the decoding process and emits a response, which could be regarded as a query expansion problem in language generation. In view of the prior success of transformer-based language models in expanding a concept to a descriptive sentence \cite{DBLPabs02610, jietal2020language, shwartznsupervised} and the state-of-the-art performance of pre-trained language generation models, training a language generation model from scratch is unnecessary. It is also worth mentioning that our innovation mainly lies in learning expressive KGs and conversation representations via mutual information maximization as well as the explicit KG reasoning. Thus, we adopt GPT-2 \cite{DBLgpt22610} in KECR to generate the response for the next turn $y_{t+1}=\{c_1, c_2, \cdots, c_{N}\}$ and it consists of $N$ tokens:
$c_n = \underset{\forall c}{\mathrm{argmax}}\, P_{\theta_{GPT-2}}(c\mid c_{<n}, \mu)$, where $\mu$ denotes the two entities resulted from the reasoning process earlier, and $\theta_{GPT-2}$ denotes the pre-trained GPT-2 model that parameterizes the probability calculation of token $c$. When generating the $n$-th token in the response, GPT-2 encodes the concatenated embeddings of selected entities $\mu$ and the context $x_t$, then decode tokens inside the response sequentially with nucleus sampling method \cite{Holtzman2020The, budzianowvu9ello}, which helps to generate semantically richer responses.

\subsection{Training}
The training process of our proposed KECR contains two phases: (1) pre-training for the static context information modelling; and (2) joint training for the dialogue policy agent and dynamic reasoning module.
For the first phase, the training objective is to fuse the mutual information from the graph embedding and the context embedding via mutual information estimation. We optimize the semantic space of each type of embedding by maximizing the lower bound of their mutual information defined by the BCE loss $L_{MI}$ in Section 4.1.3. For the joint training phase, we first infer the response action based on the conversation context. Then, we integrate the damping effect and the self-attention mechanism to derive the user preference embedding. Conditioned on the dialogue action, context embedding, and user preference embedding, KECR reasons on the KG to find proper neighbours to formulate the next response. KECR's model parameters are refined by minimizing the sum of $L_a$ and $L_r$. Note that GPT-2 has already been trained and optimized on a massive Web corpus, hence we directly inherit the original GPT-2 model to decode the conversational response. The detailed training process can be found in Algorithm 1.
\begin{algorithm}[t]
	\SetAlgoNoLine
	\KwIn{Knowledge graph $\mathcal{G}= (\mathcal{V}, \mathcal{R})$, training data $\mathcal{X}$}
	\KwResult{Fine tuned parameter for graph embedding $\theta_g$, context embedding $\theta_c$, policy agent $\theta_a$, reasoning module $\theta_r$  }
	Randomly initialize $\theta_g$, $\theta_c$, $\theta_a$, $\theta_r$\;
	Pre-train $\theta_g$ and $\theta_c$ by maximizing the MIM loss $L_{MI}$ in Eq.(4)\;
	\For{$t=1$ $\to$$|\mathcal{X}|$ }{
		Acquire the entity embedding $e_v$ from $\mathcal{G}$ and the context embedding $q$ from $\mathcal{X}$ by Eq.(1) and Eq.(2), respectively\;
		Predict the dialogue action $a$ by Eq.(5)\;
		Compute user preference representation $u$ by Eq.(7-9)\;
		Perform reasoning over the KG from the starting point, calculate the relevent score of neighbors by Eq.(10)\;
		Conduct second step reasoning by Eq.(10)\;
		Update $\theta_a$ and $\theta_r$ with Eq.(6,11-12)
	}\
	Return $\theta_g$, $\theta_c$, $\theta_a$, $\theta_r$
	\caption{The training algorithm for KECR.}
	\label{algorithm}
\end{algorithm}

\section{Experiments}
We evaluate KECR on two real-world datasets. Specifically, we aim to answer the following research questions (RQs):

\noindent\textbf{RQ1:} How does our proposed KECR perform compared with state-of-the-art conversational recommendation methods?

\noindent\textbf{RQ2:} Are our mutual information estimation and preference reasoning modules really effective?

\noindent\textbf{RQ3:} How can KECR smoothen the conversation flow and provide persuading recommendations?

\subsection{Datasets}
We conduct experiments on two publicly available datasets, REDIAL \cite{DBLPabs181207617} and GoRecDial \cite{kangrecommendation}, which were collected with Wizard-of-Oz paradigm \cite{10357417.357420}. The conversations are generated by human volunteers recruited via Amazon Mechanical Turk. The oringinal corpus contains many grammatical errors and incomplete tokens, so we manually corrected them before the training. Both datasets focus on the movie recommendation domain. REDIAL involves the free chats between the wizard and the seeker. The wizard captures the user preference and selects a suitable recommendation for the seeker. At least four different movies are mentioned in a single conversation. GoRecDial additionally provides the watching history of the seeker. The wizard chats with the seeker to acquire the seeker's preference, and recommends a movie from five given candidates.

To fully simulate realistic application scenarios, we construct KG by mapping movies to DBPedia via exact name matching, and filter out inappropriate relations. Finally, the relation of \textit{Actor}, \textit{Director}, \textit{Time}, \textit{Genre}, \textit{Subject} are retained from the DBPedia. Apart from these relations, we append the inverse relation in our KG to enhance the graph embedding of the attribute entities as mentioned in Session \uppercase\expandafter{\romannumeral4}.A.(1), enabling item and attribute entities to have enough neighbours for embedding learning. We also create a \textit{Belong} relation, that connects attributes with general clarifying entities defined by relations. Table \uppercase\expandafter{\romannumeral2} illustrates the statistics of the two datasets along with their KGs after preprocessing.

\begin{table}[]
\renewcommand{\arraystretch}{1.1}
\centering
	\caption{Statistics of REDIAL and GoRecDial.}
	\begin{tabular}{c|c|c|c}
		\hline
		\multicolumn{2}{c|}{}                                & REDIAL & GoRecDial \\ \hline
		\multirow{3}{*}{Conversations} & Interactions       & 10006         & 9125       \\ \cline{2-4}
		& \#Utterances       & 182150         & 170904       \\ \cline{2-4}
		& \#Movies     & 6924       & 3782      \\ \cline{2-4} \hline
		\multirow{3}{*}{KG}        & \#Entity      & 30471        & 19308       \\ \cline{2-4}
		& \#Relation    & 12           & 12          \\ \cline{2-4}
		& \#Triple      & 392682       & 227384     \\ \hline
	\end{tabular}
\end{table}

\subsection{Experimental Settings}
\subsubsection{Implementation Details}
We follow the procedure introduced in Session \uppercase\expandafter{\romannumeral4}.D to implement our system with PyTorch. We randomly split each dataset for training, validation and test with the ratio of $8:1:1$. For convenience, the graph and the context embeddings share the same dimension size of 128. The embedding layer of the RGCN $L$ is set to 1, and the normalization factor $Z_{v,r}$ is set to 1. The damping factor is set to 0.95 in Eq.(9). We adopt a learning rate of 0.001, and the batch sizes are respectively 10 and 30 for pre-training and joint training. Adam is adopted as the optimizer with a weight decay of 0.01. The pre-trained language model BERT and GPT-2 used in KECR are the basic types that are published in Transformers toolkit\footnote{https://github.com/huggingface/transformers}. KECR did not involve fine-tuning of the pre-trained language model. Therefore, we freeze the parameters of BERT and GPT-2 during the training process.

\subsubsection{Baselines}
Research on CRS has emerged in recent years, which explores the application in various scenarios and settings. To verify the performance of our proposed KECR, we select several state-of-the-art baseline methods designed for the same application setting as our proposed method. The baselines are summarized as follows:
\begin{itemize}
\item ReDial: This is a benchmark model for REDIAL \cite{DBLPabs181207617}. It consists of an RNN-based language generation module, a CF-based recommender and a sentiment analysis module.

\item KBRD:
This is a knowledge-based CRS \cite{chental2019towards} that allows mutual enhancement between the knowledge of user preferences and the generated recommendation results.

\item KGSF:
This method tackles the semantic gap problem during the conversational recommendation process \cite{10114533944863403143}. Based on the knowledge-enhanced embedding of user preference, the recommender component generates suitable recommendations and feeds them to a transformer-based decoder to generate responses.

\item CRWalker:
This method transforms the reasoning process of user preference during the conversational recommendation as a tree-based walk on the KG \cite{DBLPabs201010333}. The model expands the user preference reasoning tree from the root to the end of the leaves. The decoder integrates the reasoning tree to formulate the responses at each conversation turn.
\end{itemize}
Although there are other recent CRSs \cite{1011453404833462839, 10404835462913, 1061913371769, wenqiangleikdd20}, they are inapplicable for comparison due to different task settings. For example, \cite{wenqiangleikdd20, 1061913371769} are designed to formulate a series of yes/no questions based on the preference estimation, and \cite{10404835462913} is designed to learn the dialogue policy during conversational recommendation with reinforcement learning.

\begin{table}[]
	\centering
	\renewcommand{\arraystretch}{1.1}
	\setlength\tabcolsep{1.3pt}
	\caption{Recommendation performance comparison of all methods on two datasets, where the best performance is boldfaced. Improvements over all baselines are statistically significant with $p<0.01$.}
	\begin{tabular}{c|c|c|c|c|c}
		\hline
		\multirow{2}{*}{} & \multicolumn{3}{c|}{REDIAL} & \multicolumn{2}{c}{GoRecDial} \\ \cline{2-6}
		&R@1  (\%)	&R@10 (\%) & Impr. (R@1) &R@1  (\%) & Impr. (R@1) \\ \hline
		ReDial&1.93 &10.30 & +102.59\% &69.18 & +14.21\% \\ \hline
		KBRD&3.02  &15.73 & +29.47\% &74.52 & +6.03\% \\ \hline
		KGSF&3.78 &16.15 & +3.44\% &77.36 & +2.13\% \\ \hline
		CRWalker&2.05 &11.42 & +90.73\% &72.83 &	 +8.49\% \\ \hline
		KECR&$\mathbf{3.91}$  &$\mathbf{16.49}$ & - &$\mathbf{79.01}$ & -\\ \hline
	\end{tabular}
	\vspace{-0.2cm}
\end{table}

\begin{table}[]
\renewcommand{\arraystretch}{1.1}
	\centering
	\caption{Response generation performance comparison of all methods on two datasets by Distinct-n, where the best performance is boldfaced. We additionally underline the second-best performance on Distinct-n.}
	\begin{tabular}{c|c|c|c|c|c|c}
		\hline
		\multirow{2}{*}{} & \multicolumn{3}{c|}{REDIAL} & \multicolumn{3}{c}{GoRecDial} \\ \cline{2-7}
		&Dist-2 &Dist-3	&Dist-4		
		&Dist-2	&Dist-3	&Dist-4	           \\ \hline
		ReDial &10.61 &13.07 &12.40 &11.66  &13.23  &12.61	\\ \hline
		KBRD&11.38 &23.44 &30.21  &13.46  &17.19 &24.92 	\\ \hline
		KGSF &13.45 &27.53 &34.51 &14.14  &18.66 &27.59 	\\ \hline
		CRWalker &$\mathbf{27.24}$ &$\mathbf{39.61}$ &$\mathbf{49.57}$ & $\mathbf{22.63}$ &$\mathbf{32.36}$ & $\mathbf{43.12}$	\\ \hline
		KECR &$\underline{\mathbf{26.45}}$ &$\underline{\mathbf{36.10}}$ &$\underline{\mathbf{42.18}}$ &$\underline{\mathbf{18.13}}$  &$\underline{\mathbf{27.62}}$ &$\underline{\mathbf{39.15}}$ 	\\ \hline
	\end{tabular}
	\vspace{-0.2cm}
\end{table}

\subsubsection{Evaluation Metric}
We evaluate the performance of different methods in two ways, automatic evaluation and human evaluation. The automatic evaluation includes the assessment of recommendation accuracy and the quality of language generation. Following \cite{10114533944863403143}, we adopt Recall@k (R@k) to evaluate the recommendation performance, Distinct n-gram (Dist-n) to measure the diversity of generated responses, and BLEU to compare the similarity between the generated responses and the ground truth sentences. Different from the previous methods that measure the performance on sentence-level, we evaluate the performance on context-level, which is more comprehensive for the conversation scenario.

The automatic evaluation metrics sometimes may not reflect the actual property of the generated responses \cite{novikova-etal-2017-need}. The similarity-based evaluation metric (i.e. BLEU) may cheat the decoder to generate responses containing more similar tokens, ignoring the grammar and fluency. Therefore, we follow \cite{liuconversational} to invite external human volunteers to give their ratings (ranging from 1-5) on 4 properties of the generated responses, \textit{fluency}, \textit{coherence}, \textit{informativeness} and \textit{syntax}.

\subsection{Performance Comparison}
Table \uppercase\expandafter{\romannumeral3} and Table \uppercase\expandafter{\romannumeral4} present the experimental results of our proposed method as well as the baselines\footnote{As mentioned in Session \uppercase\expandafter{\romannumeral5}.A, the GoRecDial provides 5 recommendation candidates during the conversation, so we only evaluate R@1 on the GoRecDial.}. Obviously, our proposed method KECR achieves the best recommendation performance and the second-best performance on response diversity.

\textbf{Recommendation Effectiveness.} As shown in Table \uppercase\expandafter{\romannumeral3}, most KG-based methods achieve better recommendation performance compared with CF-based ReDial. That is mainly because ReDial generates the recommendation solely relies on the collaborative signal in the training corpus. The use of plain CF makes ReDial unable to incorporate auxiliary information (e.g., KG) to distill the user preference, hindering it from generating accurate recommendations, especially in the sparse conversation scenario. KBRD provides an effective method to incorporate the KG as complementary information to enrich the item embeddings from RGCN. It indicates the effectiveness of using KGs to improve CRSs. KGSF utilizes two KGs, i.e., ConceptNet and DBPedia, to fuse the token-based and item-based semantics, so as to facilitate fine-grained user preference modeling and recommendation. However, both KG-based methods only focus on enriching the item/token embeddings during the conversation. They cannot utilize the explicit KG connections to reason the user preference and narrow down the recommendation candidates. CRWalker introduces a reasoning module to generate the recommendation, but it ignores the context information that can help optimize the reasoning process, resulting in inferior recommendation performance.

\textbf{Dialogue Quality.} From the perspective of response generation, Table \uppercase\expandafter{\romannumeral4} gives a fair comparison of all methods, where ReDial still makes the worst generation performance. ReDial generates the response with the RNN-based encoder-decoder architecture, which strictly follows the language rules learned from the training corpus. The lack of auxiliary information from KG makes it unable to fully distill the semantics from the conversation context, failing to generate high-quality responses. KBRD and KGSF introduce the vocabulary bias from KG into a transformer module to enrich the response representation. However, as mentioned above, their methods mainly focus on enriching the item/token embeddings during the conversation. They cannot utilize the explicit KG connections to justify and explain their recommendations. CRWalker achieves the highest score on Dist-n. However, this is mainly because their inaccurate predictions of recommendation will further mislead the subsequent reasoning process and bring irrelevant vocabulary biases to contaminate the decoding process, which leads to abnormally large Dist-n scores.

Our proposed KECR addresses all aforementioned drawbacks of baselines and achieves superior performance on both Dist-n and Recall via explicit reasoning on the KG. The main goal for a CRS is to recommend suitable items for users, which is evaluated by R@K, where KECR achieves the best performance compared with baselines. The reasoning process could identify informative paths on the KG to find suitable recommendations when scarce preference information could be accessed. Beyond the recommendation accuracy, we want the system to generate diverse responses instead of similar responses as in the training corpus. The reasoning component in KECR could provide a rational vocabulary diversity to expand the actual semantic representations, and provide diversification of the language generation. These extra vocabulary biases are sequentially injected into the large pretrained language model, enabling our system to generate faithful explanations based on user preferences.

\subsection{Ablation Study}
The static context information modelling and the dynamic reasoning flow are the two main components of our proposed KECR. We especially introduce the mutual information estimation and dynamic reasoning module in each part to enhance the performance of the basic CRS. To verify the effectiveness of newly proposed components in our system, we design two variants of our KECR, \textbf{KECR-M} and \textbf{KECR-R}. KECR-M keeps all other components but removes mutual information estimation. KECR-R maintains the mutual information estimation to fuse the two channel embedding signals but makes changes on the reasoning part. We replace the reasoning component with a random selection module that randomly selects one neighbour of the start point to formulate the next response. It also conducts two-step reasoning to generate a pseudo explanation for this selection. To demonstrate the performance variation, we utilize BLEU and Dist-3 for dialogue quality and R@10/R@1 (on REDIAL/GoRecDial) for recommendation.
\begin{table}[t]
\renewcommand{\arraystretch}{1.1}
	\centering
	\caption{Performance comparison of all ablation methods on two datasets by Recall@10, BLEU and Distinct-3, where the best performance is boldfaced.}
	\begin{tabular}{c|c|c|c|c}
		\hline
		\multirow{2}{*}{} & \multicolumn{2}{c|}{REDIAL} & \multicolumn{2}{c}{GoRecDial} \\ \cline{2-5}
		&R@10 (\%)	 &Dist-3			
		&R@1 (\%)			&Dist-3		           \\ \hline
		KECR&$\mathbf{16.49}$  &36.10 &$\mathbf{79.01}$  &27.62 	\\ \hline
		KECR-M&15.63 &37.18 &76.52 & 28.46   	\\ \hline
		KECR-R&7.36  &$\mathbf{45.64}$ &43.82  &$\mathbf{48.6}$ 	\\ \hline
	\end{tabular}
	\vspace{-0.3cm}
\end{table}

\begin{table}[]
\renewcommand{\arraystretch}{1.1}
	\caption{Human evaluation of all methods on sampled conversation on fluency, coherence, informativenss and syntax, where the best performance is boldfaced.}
	\begin{tabular}{c|c|c|c|c}
		\hline
		&Fluency & Coherence & Informativeness & Syntax  \\ \hline
		ReDial&4.1 &4.2  &3.2 &3.5  	\\ \hline
		KBRD&4.2 &4.0  &4.1 &4.1  	\\ \hline
		KGSF&4.1 &4.0  &4.2 &4.1  	\\ \hline
		CRWalker &4.0	&4.3	&4.3 	&4.4	\\ \hline
		KECR &$\mathbf{4.4}$	&$\mathbf{4.6}$	&$\mathbf{4.7}$	&$\mathbf{4.5}$	\\ \hline
	\end{tabular}
	\vspace{-0.5cm}
\end{table}

From Table \uppercase\expandafter{\romannumeral5}, we can observe that the default setting of our proposed method still keeps the best performance over these two variants. The improvement of the score on R@K for default setting represents both the mutual information estimation and the reasoning module could provide a better guideline for finding the target items. KECR-M shows constant performance degeneration compared with the default one, demonstrating the effect of the mutual information estimation. We also notice that KECR-R has achieved a higher score on Dist-n compared with the default one. That is mainly because the random selection policy introduces false tokens into the final decoding stage, which would increase the similarity-based Dist-n score. The improvement on Dist-n score also comes with the performance damping on BLEU and R@K. KECR-R cannot accurately identify the user preference and recommend suitable items to the user, failing in the primary task of a conversation recommendation.

\subsection{Human Evaluation}
The automatic evaluation metrics sometimes overly rely on the similarity of the n-gram features, which cannot comprehensively evaluate the generated response's actual quality, e.g., fluency, coherence, informativeness and syntax. However, real-world applications care more about the generated sentence's actual quality than various evaluation metrics. Therefore, we follow \cite{liuconversational} and recruit ten external volunteers to contribute their ratings (ranging from 1-5) on the generated conversations from 4 perspectives: fluency, coherence, informativeness and syntax. A higher score means the volunteers have a higher agreement on one specific perspective of language quality. We sampled 20 conversations from REDIAL to interact with our proposed method and different baselines.

Table \uppercase\expandafter{\romannumeral6} presents the collected blind reviews from the human volunteers. KGSF, KBRD and CRWalker obtain a satisfying score on four evaluation aspects, mainly because of the adoption of the transformer-based language model and the auxiliary information from the KG. ReDial inherits the property of the RNN-based language models that tends to imitate the conversation behaviour appearing in the training corpus when decoding, which could achieve a good performance on  fluency and coherence. Finally, our proposed KECR consistently performs better than all baselines. KECR could make reasoning over the KG to provide informative entities as guidances to enhance the coherence and the informativeness of the generated response. Also, it gets benefits from the powerful pre-trained language model, which is trained on massive open web text. The pre-trained language model expands the key queries obtained from the reasoning process to formulate the response, proving the fluency and the syntax of the generated responses.

\begin{figure}
	\centering
	\includegraphics[width=0.7\linewidth]{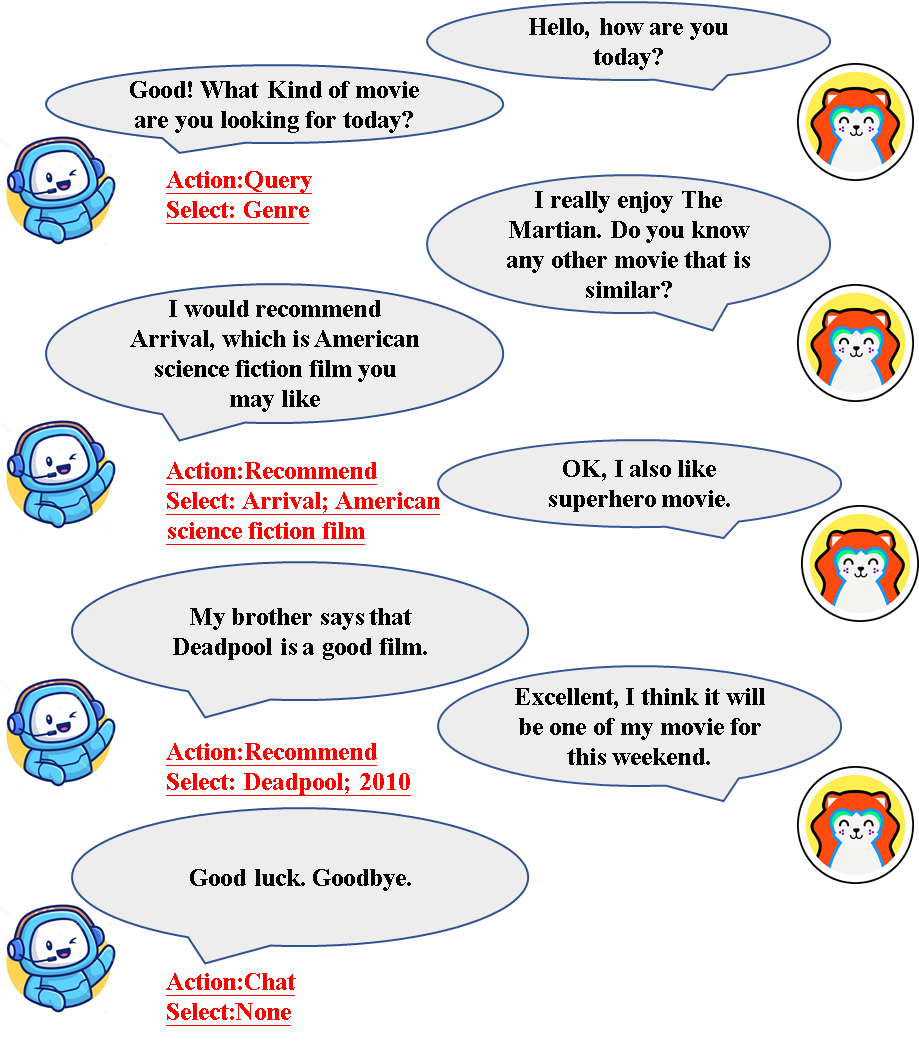}
	\caption{A sample conversation between our KECR (wizard) and a real user (seeker). The words in red font denotes the dialogue action and the selected entities during the reasoning.}
	\label{fig:demo}
\end{figure}

\subsection{Qualitative Analysis}
In this part, we present a qualitative study to illustrate how KECR is able to improve the quality of generated responses and generate explainable recommendations with persuasive descriptions.

Figure 3 presents a conversation between our human volunteer and our proposed system KECR. The human volunteer interacts with the system to ask for movie recommendations for the weekend night. The conversation starts with the greeting from the volunteer to activate the service. KECR first identifies the dialogue action as ``Query'' and selects ``Genre'' to ask for the user preference. Then, the user sequentially expressed his/her preference either with the item entity ``The Martian'' and the attribute entity ``Superhero''. The dialogue action generated from the policy agent for both conversation turns are ``Recommendation''. KECR starts to make reasoning on the KG to find a proper movie that shares the same property for the recommendation, and a corresponding reason (i.e. the second-step reasoning process). Conditioned on the selected entities, a human-like persuasive response is generated from our GPT-2 based decoder to recommend the item. Finally, there is no newly mentioned entity appearing in the conversation and the user accepts the recommendation. KECR turns to ``Chat'' action and says ``Goodbye'' to the user.

There is an interesting response during the conversation, that KECR introduces ``My brother'' in the third turn of conversation to strengthen the confidence of the recommendation. That is a side benefit from the pretrained language model that makes our system easier to generate human-like responses.

\section{Conclusions and Future Work}
In this paper, we propose a novel knowledge-enhanced conversational reasoning approach for CRS. We firstly utilize auxiliary structured knowledge from the KG to bridge the semantic gap between the attribute- and item-level representations, and adopt mutual information maximization to align the semantic space of the context and the knowledge graph. Conditioned on the aligned semantic representations, we make explicit reasoning over the KG to find a prominent reasoning chain to enhance the recommendation and the conversation generation performance. Extensive experiments on two real-world datasets demonstrate the superiority of our proposed system over state-of-the-art baselines.

As for future work, we will consider introducing multi-modal external resources (e.g., videos, demographics) to enrich the learned representations of CRS. Apart from this, we will explore the implementation of our system in real-world conversation scenarios with more conversation behaviours (e.g., chit-chat, ticket booking, movie arrangement) to fulfil the diverse demand of the user. Finally, lowering the computational cost while protecting user privacy will be a highly relevant topic when deploying CRSs on edge devices.



\end{document}